\documentclass[vecphys]{svmult}

\usepackage{makeidx}         
\usepackage{graphicx}        
\usepackage{multicol}        
\usepackage{cite}            
\usepackage[bottom]{footmisc}
\usepackage{amssymb}
\usepackage{amsbsy}

\makeindex             

\begin{document}
\title{Negative electron compressibility in the Hubbard Model}

\titlerunning{Negative electron compressibility in the Hubbard Model}

\author{A. Sherman}

\institute{Institute of Physics, University of Tartu, W. Ostwaldi Str 1, 50411 Tartu, Estonia\\ \texttt{alekseis@ut.ee}}

\authorrunning{Sherman}

\maketitle

\begin{abstract}
Using the strong coupling diagram technique, we study the one-band repulsive Hubbard model on a two-dimensional square lattice in a wide range of chemical potentials $\mu$. Infinite sequences of diagrams describing interactions of electrons with spin and charge fluctuations are taken into account. At low temperatures, regions of the negative electron compressibility are found for $\mu<0$ and $\mu>U$, where $U$ is the Hubbard repulsion. For even lower temperatures, in the neighborhood of these regions, states with the phase separation are revealed.
\end{abstract}

\section{Introduction}
The electron compressibility is defined as $\kappa=\bar{n}^{-2}({\rm d}\bar{n}/{\rm d}\mu)$, where $\bar{n}$ is the electron concentration and $\mu$ the chemical potential. Positive values of this quantity correspond to the common situation, in which an increase of the concentration leads to a growth of the chemical potential. However, cases of negative $\kappa$ are also possible, when an increase of $\bar{n}$ is accompanied with a counterintuitive {\em decrease} of $\mu$. It is a clear indication of non-rigid bands that is the characteristic property of strong electron correlations. Usually, the negative electron compressibility (NEC) is associated with thermodynamic instability and phase separation (see, e.g., \cite{Dezi}). This happens when the dependence $\bar{n}(\mu)$ is multivalued in some range of $\mu$ that is an indication of a mixture of two phases. However, there are also cases, in which NEC does not lead to phase separation (see, e.g., \cite{Riley}), and one of these cases will be considered below. Most frequently NEC is observed in low-dimensional dilute electron systems \cite{Dezi,Riley,Kravchenko,Eisenstein}. Recently NEC was also detected in the quasi-three-dimensional crystal (Sr$_{1-x}$La$_x$)$_3$Ir$_2$O$_7$ at higher electron concentrations \cite{He}.

In the one-band repulsive Hubbard model on a two-dimensional (2D) square lattice, NEC is observed already in the Hubbard-I approximation \cite{Hubbard63,Hubbard64} as soon as $\mu$ leaves the usually considered range $0<\mu< U$ and the temperature $T$ is low enough. Here $U$ is the on-site Coulomb repulsion. The origin of the NEC is the discreteness of the energy levels in the Hubbard atom and their crossing, which occurs when $\mu$ traverses the energies 0 and $U$. Due to occupation-probability factors the first-order cumulant of electron operators, which determines Green's function in the Hubbard-I approximation, changes drastically in the ranges $-T\lesssim\mu\lesssim T$ and $U-T\lesssim\mu\lesssim U+T$. This alteration is connected with the change of the lowest state -- for $\mu\ll -T$ and $\mu\gg U+T$ it is empty and doubly occupied states, while for $T\ll\mu\ll U-T$ this state is singly occupied. For $U\gg t$, where $t$ is the hopping constant, the former two states are associated with weakly correlated bands of the Hubbard model, whereas the latter state is responsible for the case of strong correlations. Dependencies $\bar{n}(\mu)$ in these three cases are markedly different that leads to the NEC in transition regions $\mu\approx 0$ and $\mu\approx U$.

Of course, the Hubbard-I approximation gives an oversimplified picture. It is necessary to prove whether these NEC regions survive if charge and spin fluctuations are taken into account. This question is considered in the present work. To answer it we use the strong coupling diagram technique (SCDT) \cite{Vladimir,Metzner,Pairault,Sherman16,Sherman18,Sherman19a,Sherman19b}. In this approach, Green's functions are calculated using series expansions in powers of the electron kinetic energy. Terms of the series are products of hopping constants and on-site cumulants of electron operators. The linked-cluster theorem is valid and partial summations are allowed in the SCDT (concise description of the approach can be found in \cite{Sherman16}). In SCDT, the interaction of electrons with charge and spin fluctuations is described by diagrams with ladder inserts. If these ladders are constructed from renormalized hopping lines and second-order cumulants, ladders of all lengths can be summed \cite{Sherman18}. As a consequence, fluctuations of all ranges are taken into account in an infinite crystal. In \cite{Sherman18,Sherman19a}, it was shown that the spectral function, magnetic susceptibility, double occupancy and squared site spin calculated in the SCDT are in satisfactory agreement with results of exact diagonalizations, Monte Carlo simulations, numerical linked-cluster expansions and experiments with ultracold fermionic atoms in 2D optical lattices in wide ranges of repulsions, temperatures and concentrations. Below this approach is applied for the consideration of the influence of charge and spin fluctuations on the NEC. We find that the fluctuations do not destroy it and the NEC is observed at low temperatures. To attain even lower $T$, we reduce the set of considered diagrams. In this approximation, regions of bistability or phase separation are found near the NEC domains.

\section{Model and SCDT method}
The fermionic 2D one-band Hubbard model \cite{Hubbard63,Hubbard64} is used in this work. Its Hamiltonian reads
\begin{equation}\label{Hamiltonian}
H=\sum_{{\bf ll'}\sigma}t_{\bf ll'}a^\dagger_{{\bf l'}\sigma}a_{{\bf l}\sigma}
+\frac{U}{2}\sum_{{\bf l}\sigma}n_{{\bf l}\sigma}n_{{\bf l},-\sigma},
\end{equation}
where 2D vectors ${\bf l}$ and ${\bf l'}$ label sites of a square plane lattice, $\sigma=\uparrow,\downarrow$ is the spin projection, $a^\dagger_{{\bf l}\sigma}$ and $a_{{\bf l}\sigma}$ are electron creation and annihilation operators, $t_{\bf ll'}$ is the hopping constant and $n_{{\bf l}\sigma}=a^\dagger_{{\bf l}\sigma}a_{{\bf l}\sigma}$. In this work, only the nearest neighbor hopping constant $t$ is taken to be nonzero.

In this work, we calculate the dependence $\bar{n}(\mu)$ for the Hamiltonian (\ref{Hamiltonian}). Here $\bar{n}=2\langle n_{{\bf l}\sigma}\rangle$, angular brackets denote the statistical averaging with the operator ${\cal H}=H-\mu\sum_{{\bf l}\sigma} n_{{\bf l}\sigma}$ and the factor 2 is connected with two spin states. To find this dependence the one-particle Green's function
\begin{equation}\label{Gfl}
G({\bf l}',\tau';{\bf l},\tau)=\langle{\cal T}\bar{a}_{{\bf l}'\sigma}(\tau')a_{{\bf l}\sigma}(\tau) \rangle
\end{equation}
is calculated using the SCDT. Here ${\cal T}$ is the chronological operator arranging operators from right to left in ascending order of times $\tau$ and time dependencies are determined by ${\cal H}$,
$$\bar{a}_{{\bf l}\sigma}(\tau)={\rm e}^{{\cal H}\tau}a^\dagger_{{\bf l}\sigma} {\rm e}^{-{\cal H}\tau}.$$

\begin{figure}[t]
\centering
\includegraphics[width=10 cm]{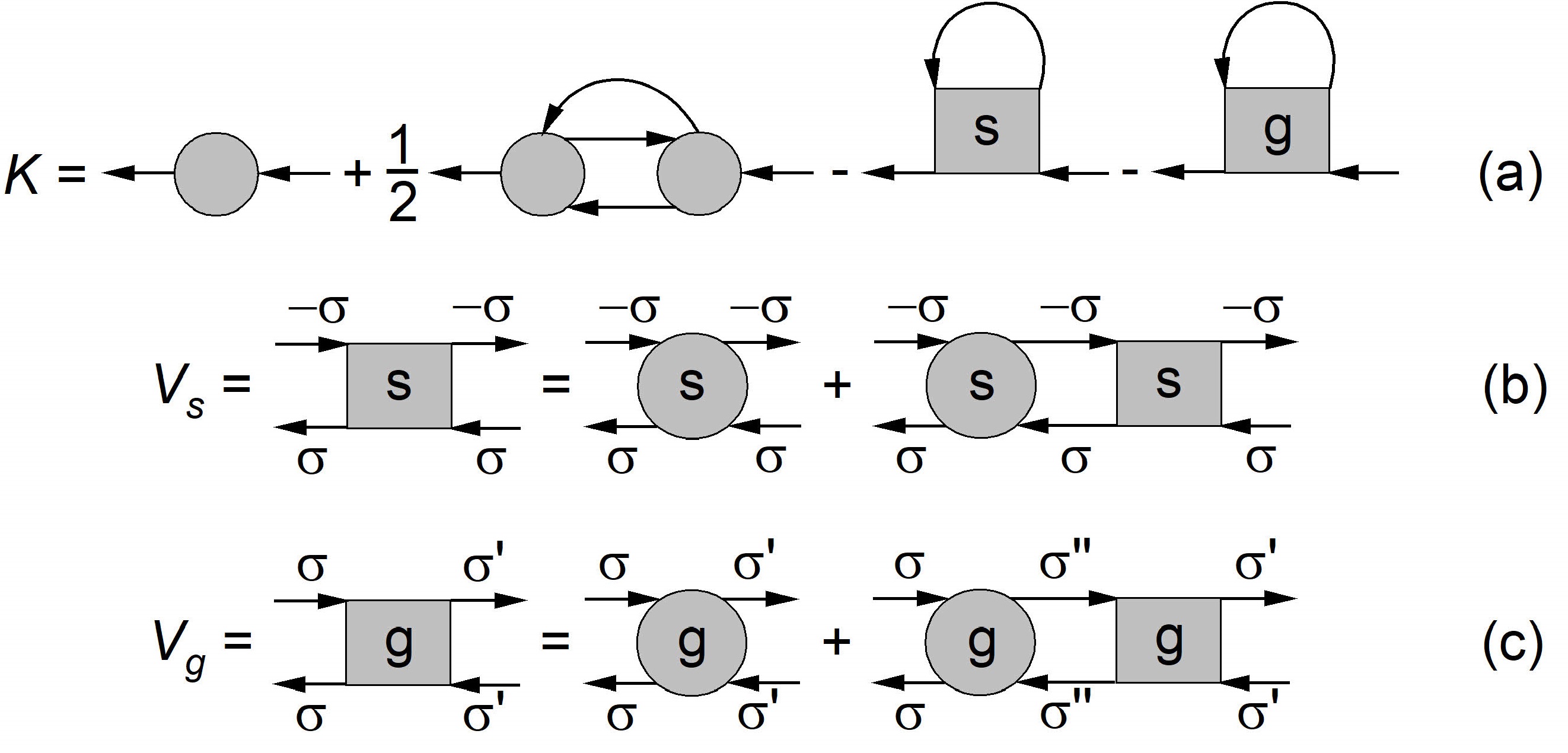}
\caption{(a) Diagrams in $K({\bf k},j)$ taken into account in the present work. (b,c) Bethe-Salpeter equations for the four-leg diagrams in part (a).}\label{Fig1}
\end{figure}

As in the weak coupling diagram technique \cite{Mahan}, diagrams of the SCDT describing Green's function (\ref{Gfl}) can be classed into reducible and irreducible. The latter cannot be divided into two disconnected parts by cutting a hopping line. The sum of all such diagrams is termed the irreducible part $K$. The Fourier transform of Green's function is expressed through this quantity as
\begin{equation}\label{Larkin}
G({\bf k},j)=\left\{\left[K({\bf k},j)\right]^{-1}-t_{\bf k}\right\}^{-1},
\end{equation}
where {\bf k} is a 2D wave vector, $j$ is an integer defining the Matsubara frequency $\omega_j=(2j-1)\pi T$ and $t_{\bf k}$ is the Fourier transform of $t_{\bf ll'}$. In the considered case, diagrams in the particle-hole channel make the main contribution to $K$ \cite{Bychkov,Bickers}. These diagrams are shown in Figure~\ref{Fig1}.
In this figure, circles without letters are cumulants of the first and second orders
\begin{eqnarray*}
&&C_1(\tau',\tau)=\big\langle{\cal T}\bar{a}_{{\bf l}\sigma}(\tau')a_{{\bf l}\sigma}(\tau)\big\rangle_0\,,\\
&&C_2(\tau_1,\sigma_1;\tau_2,\sigma_2;\tau_3,\sigma_3;\tau_4,\sigma_4)
=\big\langle{\cal T}\bar{a}_{{\bf l}\sigma_1}(\tau_1)a_{{\bf l}\sigma_2}(\tau_2) \bar{a}_{{\bf l}\sigma_3}(\tau_3)a_{{\bf l}\sigma_4}(\tau_4)\big\rangle_0\\
&&\quad\quad-\big\langle{\cal T}\bar{a}_{{\bf l}\sigma_1}(\tau_1)a_{{\bf l}\sigma_2}(\tau_2)\big\rangle_0\big\langle{\cal T}\bar{a}_{{\bf l}\sigma_3}(\tau_3)a_{{\bf l}\sigma_4}(\tau_4)\big\rangle_0
+\big\langle{\cal T}\bar{a}_{{\bf l}\sigma_1}(\tau_1)a_{{\bf l}\sigma_4}(\tau_4)\big\rangle_0\big\langle{\cal T}\bar{a}_{{\bf l}\sigma_3}(\tau_3)a_{{\bf l}\sigma_2}(\tau_2)\big\rangle_0\,,
\end{eqnarray*}
where the subscript 0 near brackets indicates that time dependencies and averages are determined by the Hamiltonian of the Hubbard atom
\begin{equation}\label{Hubbard}
{\cal H}_{\bf l}=\sum_\sigma\left[(U/2)n_{{\bf l}\sigma}n_{{\bf l},-\sigma}- \mu n_{{\bf l}\sigma}\right].
\end{equation}
Due to the translation symmetry, the cumulants are identical on all lattice sites. The last two terms in $K({\bf k},j)$ correspond to diagrams with ladder inserts and describe the interaction of electrons with charge and spin fluctuations. The vertices $V_s$ and $V_g$ satisfy the Bethe-Salpeter equations (BSE) shown in Figures~\ref{Fig1}(b) and (c), in which circles with letters are irreducible four-leg diagrams which cannot be divided into two disconnected parts by cutting a pair of horizontal oppositely directed hopping lines. Thanks to the possibility of the partial summation, bare internal lines $t_{\bf k}$ in Figure~\ref{Fig1} can be transformed into dressed ones,
\begin{equation}\label{hopping}
\theta({\bf k},j)=t_{\bf k}+t^2_{\bf k}G({\bf k},j).
\end{equation}

In this work, the irreducible four-leg diagrams in Figures~\ref{Fig1}(b) and (c) are approximated by the respective second-order cumulants. As mentioned in the Introduction, this approach gave results in a satisfactory agreement with reliable calculation and experimental methods, at least in the range of chemical potentials $T\ll\mu$, $T\ll U-\mu$. Besides, magnetic susceptibilities obtained in this approximation were compared with those calculated in a more complex approach, in which the longitudinally irreducible vertices in Figures~\ref{Fig1}(b) and (c) were approximated by full sums of transversal ladders constructed from dressed hopping lines and second-order cumulants. The susceptibilities appeared to be close in these two approaches in the mentioned range of $\mu$. This gives grounds to use the simpler approximation, with second-order cumulants as irreducible vertices, for investigating a wider range of chemical potentials. In this case, $V_s$ and $V_g$, apart from frequencies, depend only on the transfer momentum. The irreducible part and BSEs read
\begin{eqnarray}\label{K}
K({\bf k},j)&=&C_1(j)-\frac{T}{N}\sum_{{\bf k'}j'}\theta({\bf k'},j')\big[V_{s,\bf k- k'}(j,\sigma;j,\sigma;j',-\sigma;j',-\sigma)
+V_{g,\bf k-k'}(j,\sigma;j,\sigma;j',\sigma;j',\sigma)\big]\nonumber\\
&&+\frac{T^2}{2N^2}\sum_{{\bf k'}j'\nu}\theta({\bf k'},j'){\cal T}_{\bf k-k'}(j+\nu,j'+\nu)\nonumber\\
&&\quad\times\Big[C_2(j,\sigma;j+\nu,\sigma;j'+\nu,-\sigma;j',-\sigma)
C_2(j+\nu,\sigma;j,\sigma;j',-\sigma;j'+\nu,-\sigma)\nonumber\\
&&\quad+\sum_{\sigma'} C_2(j,\sigma;j+\nu,\sigma';j'+\nu,\sigma';j',\sigma)
C_2(j+\nu,\sigma';j,\sigma;j',\sigma;j'+\nu,\sigma')\Big],
\end{eqnarray}
\begin{eqnarray}
&&V_{s{\bf k}}(j+\nu,\sigma;j,\sigma;j',-\sigma;j'+\nu,-\sigma)
=C_2(j+\nu,\sigma;j,\sigma;j',-\sigma;j'+\nu,-\sigma)\nonumber\\
&&\quad\quad+T\sum_{\nu'}C_2(j+\nu,\sigma;j+\nu',\sigma;j'+\nu',-\sigma;j'+\nu,-\sigma)
{\cal T}_{\bf k}(j+\nu',j'+\nu')\nonumber\\
&&\quad\quad\times V_{s{\bf k}}(j+\nu',\sigma;j,\sigma;j',-\sigma;j'+\nu',-\sigma), \label{Vs}\\
&&V_{g{\bf k}}(j+\nu,\sigma';j,\sigma;j',\sigma;j'+\nu,\sigma')
=C_2(j+\nu,\sigma';j,\sigma;j',\sigma;j'+\nu,\sigma')\nonumber\\
&&\quad\quad+T\sum_{\nu'\sigma''}C_2(j+\nu,\sigma';j+\nu',\sigma'';j'+\nu',\sigma''; j'+\nu,\sigma')
{\cal T}_{\bf k}(j+\nu',j'+\nu')\nonumber\\
&&\quad\quad\times V_{g{\bf k}}(j+\nu',\sigma'';j,\sigma;j',\sigma;j'+\nu',\sigma''). \label{Vg}
\end{eqnarray}
In Equations~(\ref{K})--(\ref{Vg}), cumulants of the first and second orders are given by the equations \cite{Vladimir,Sherman06,Sherman08}
\begin{eqnarray}
&&C_1(j)=Z^{-1}\left[\left({\rm e}^{-\beta E_1}+{\rm e}^{-\beta E_0}\right)g_{01}(j)+\left({\rm e}^{-\beta E_1}+{\rm e}^{-\beta E_2}\right)g_{12}(j)\right], \label{C1}\\
&&C_2(j+\nu,\sigma';j,\sigma;j',\sigma;j'+\nu,\sigma')=\nonumber\\
&&\quad-\left[Z^{-1}{\rm e}^{-\beta E_1}\left(\delta_{jj'}\delta_{\sigma\sigma'}-\delta_{\nu 0}\right)+Z^{-2}\left({\rm e}^{-\beta (E_0+E_2)}-{\rm e}^{-2\beta E_1}\right)\left(\delta_{jj'}-\delta_{\nu 0}\delta_{\sigma\sigma'}\right)\right]\beta'F(j)F(j'+\nu) \nonumber\\
&&\quad+Z^{-1}{\rm e}^{-\beta E_0}\delta_{\sigma,-\sigma'}Ug_{01}(j)g_{01}(j'+\nu)g_{02}(j,j'+\nu)\left[g_{01}(j+\nu)+g_{01}(j') \right]\nonumber\\
&&\quad+Z^{-1}{\rm e}^{-\beta E_2}\delta_{\sigma,-\sigma'}Ug_{12}(j)g_{12}(j'+\nu)g_{02}(j,j'+\nu)\left[g_{12}(j+\nu)+g_{12}(j') \right]\nonumber\\
&&\quad-Z^{-1}{\rm e}^{-\beta E_1}\delta_{\sigma,-\sigma'}\big\{F(j'+\nu)\big[g_{01}(j)g_{01}(j') +g_{12}(j)g_{12}(j+\nu)-g_{01}(j')g_{12}(j+\nu)\big]\nonumber\\
&&\quad+F(j)\big[g_{01}(j+\nu)g_{01}(j'+\nu) +g_{12}(j')g_{12}(j'+\nu)-g_{01}(j+\nu)g_{12}(j')\big]\big\}, \label{C2}
\end{eqnarray}
where $E_0=0$, $E_1=-\mu$ and $E_2=U-2\mu$ are eigenenergies of the Hamiltonian (\ref{Hubbard}) corresponding to zero, one and two electrons on the Hubbard atom, $Z={\rm e}^{-\beta E_0}+2{\rm e}^{-\beta E_1}+{\rm e}^{-\beta E_2}$ is the respective partition function, $g_{nm}(j)=({\rm i}\omega_j+E_n-E_m)^{-1}$, $g_{02}(j,j')=({\rm i}\omega_j+{\rm i}\omega_{j'}+E_0-E_2)^{-1}$, $F(j)=g_{01}(j)-g_{12}(j)$, $\beta=1/T$ and ${\cal T}_{\bf k}(j,j')=N^{-1}\sum_{\bf k'}\theta({\bf k+k'},j)\theta({\bf k'},j')$ with $N$ the number of sites.

The above equations describe the transition to the long-range antiferromagnetic order \cite{Sherman18}. At the derivation of the equation (\ref{C2}) $\beta'=\beta$ \cite{Vladimir,Sherman06,Sherman08}. However, for such a $\beta'$ the antiferromagnetic transition occurs at a small but finite value of $T=\zeta$. In the considered 2D system with the continuous symmetry, this result is in contradiction with the Mermin-Wagner theorem \cite{Mermin}. Hence the used approximation -- with the second-order cumulants as longitudinally irreducible four-leg vertices -- overestimates somewhat spin fluctuations. We note that the above-mentioned improvement -- the use of the full sums of vertical ladders as such vertices -- gives only a slight betterment. To remedy this defect and shift the transition temperature to zero $\beta'$ in (\ref{C2}) was set equal to $1/(T+\zeta)$, and the value of $\zeta$ was determined in the case of the strongest antiferromagnetic fluctuations -- at half-filling, $\mu=U/2$. Obtained values of $\zeta$ depend on $U/t$ -- for $U\gtrsim 4t$ it equals to $0.24t$, while for $U=t$ the parameter $\zeta=0.05t$. It may be interesting to note that the value of $\zeta$ is close to $T_{\rm spin}$, which was associated in \cite{Paiva} with the maximum in the temperature dependence of the uniform susceptibility. This maximum is a manifestation of the change in the character of long-range spin correlation with temperature.

Expressions~(\ref{Larkin})--(\ref{C2}) form a closed set of equations, which can be solved by iteration for given values of $U$, $T$, $\mu$ and the considered set of hopping constants $t_{\bf ll'}$. As a starting function in this iteration, the Hubbard-I solution was used. This solution is obtained from the above formulas if the irreducible part in (\ref{Larkin}) is approximated by the first term in the right-hand side of (\ref{K}) -- the first-order cumulant $C_1(j)$. No artificial broadening was introduced in these calculations.

Expressions~(\ref{C1}) and (\ref{C2}) are considerably simplified for chemical potentials satisfying the conditions $T\ll\mu$ and $T\ll U-\mu$. For $T\ll U$ these conditions correspond to cases of half-filling and moderate doping, which are usually considered in works on the Hubbard model. For this range of $\mu$, occupation-probability factors ${\rm e}^{-\beta E_0}$ and ${\rm e}^{-\beta E_2}$ are exponentially small in comparison with ${\rm e}^{-\beta E_1}$ and terms containing them can be dropped from the formulas. As was shown in \cite{Sherman18}, this leads to the considerable simplification of the BSEs~(\ref{Vs}) and (\ref{Vg}) reducing them to small systems of linear equations. In this work, chemical potentials outside of this range are also considered. Therefore, the mentioned simplification is inapplicable. Instead of (\ref{Vg}), it is convenient to consider equations for combinations of $V_g$, which are symmetrized and antisymmetrized over spin variables. The antisymmetrized combination coincides with $V_s$, and both combinations do not depend on the sign of spin projection. BSEs for them are systems of linear equations with matrix indices $\nu$ and $\nu'$, which depend parametrically on $j$, $j'$ and {\bf k}. In our calculations, the rank of these systems can range up to 800. Therefore, to decrease the number of numeric operations the iterative solution of these systems was used. The iterations started from the respective values of $C_2$. For considered parameters 10--15 iterations were enough to achieve convergence. These iterations were performed on each step of the general iterative solution of equations~(\ref{Larkin})--(\ref{C2}). In these external iterations, convergence was achieved in 10--40 steps.

\section{Results and discussion}
\begin{figure}[t]
\centering
\includegraphics[width=10 cm]{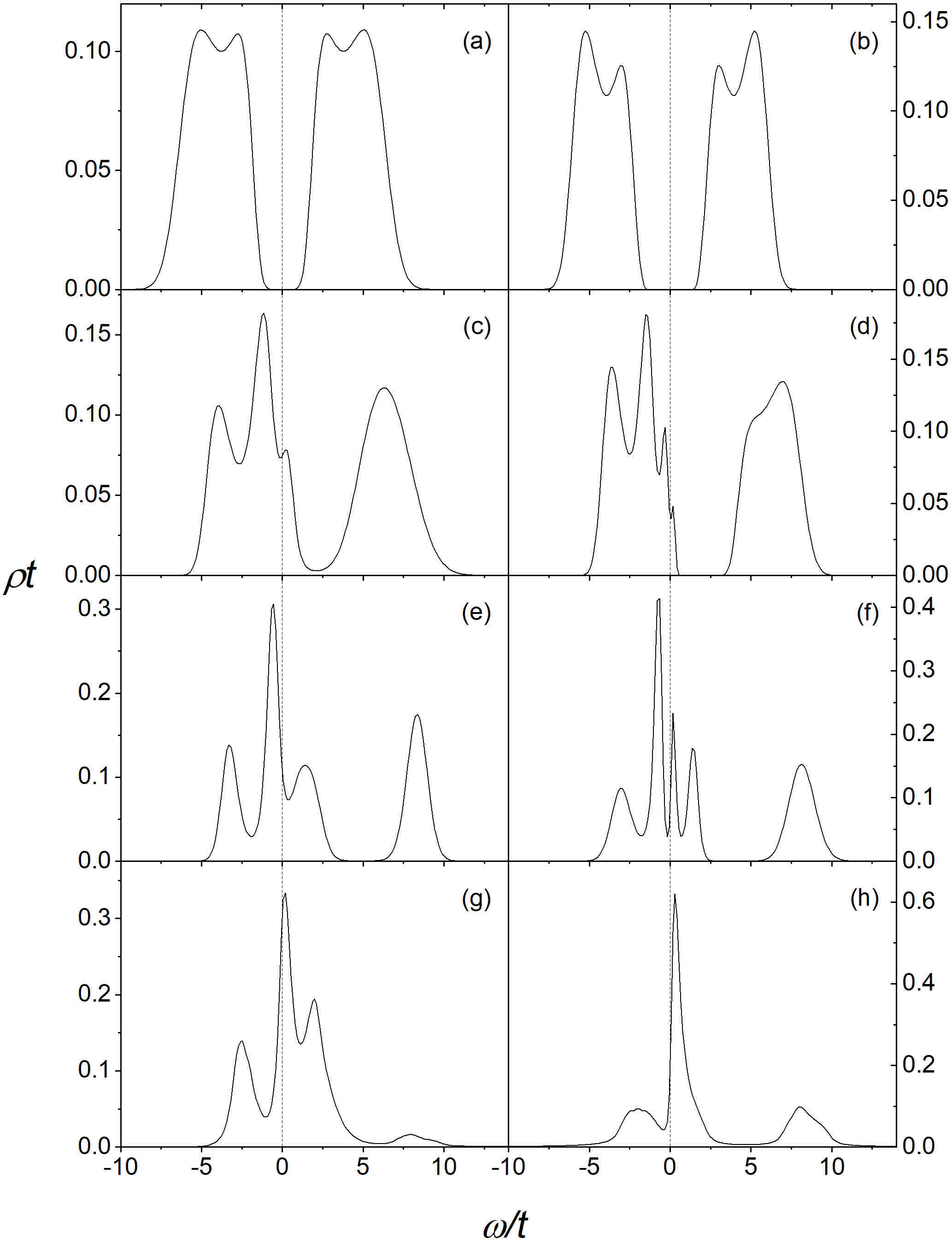}
\caption{(a) The evolution of the electron density of states with doping and temperature. $U=8t$. (a,c,e,g) $T=0.32t$; (b,d,f,h) $T=0.13t$. (a,b) $\mu=4t$, $\bar{n}=1$; (c) $\mu=2t$, $\bar{n}=0.973$; (d) $\mu=2t$, $\bar{n}=0.999$; (e) $\mu=t$, $\bar{n}=0.898$; (f) $\mu=t$, $\bar{n}=0.948$; (g) $\mu=0$, $\bar{n}=0.729$; (h) $\mu=0$, $\bar{n}=0.681$.}\label{Fig2}
\end{figure}

An example of results obtained using the above-discussed procedure is shown in Figure~\ref{Fig2}.
The figure demonstrates the temperature and filling evolution of the density of electron states (DOS) $\rho(\omega)=-(N\pi)^{-1}\sum_{\bf k}{\rm Im}G({\bf k},\omega)$. The analytic continuation from the imaginary to real frequency axis was carried out using the maximum entropy method \cite{Press,Jarrell,Habershon}. The electron concentrations were estimated using the equation $\bar{n}=(2T/N)\sum_{{\bf k}j}{\rm e}^{{\rm i}\omega_j\eta}G({\bf k},j)$ with $\eta\rightarrow +0$. Using formulas of the previous section,  $G({\bf k},j)$ was calculated for $j\leq 100$. Values of Green's function for larger $j$, which are required for evaluating $\bar{n}$ from this equation, were approximated with the asymptotics \cite{Vilk} $G({\bf k},j)\rightarrow 1/({\rm i}\omega_j)$. It was found that calculated Green's functions settle into this asymptote already at $j\ll 100$. Values of the electron concentration obtained from analytically continued $\rho(\omega)$ differ from these $\bar{n}$ by a few percent.

Due to the particle-hole symmetry of the considered model values of $\mu\leq U/2$ only will be considered below. DOSs calculated for $\mu\gg T$ are similar to those obtained earlier with a somewhat different calculation procedure \cite{Sherman18,Sherman19a,Sherman19b}. For half-filling, the Mott gap is seen around the Fermi level located at $\omega=0$ (Figures~\ref{Fig2}(a) and (b)). Suppressions of the DOS near $\omega=\pm U/2$ are connected with the strong reabsorption of electrons on the transfer frequencies of the Hubbard atom $\omega=-\mu$ and $U-\mu$ \cite{Sherman16}. With doping, for the higher temperature, a pseudogap appears around the Fermi level (Figures~\ref{Fig2}(c) and (e)). For the lower temperature, the Fermi-level peak arises inside the pseudogap (Figures~\ref{Fig2}(d) and (f)). As discussed in \cite{Sherman19b}, this peak is a manifestation of bound states of correlated electrons and mobile spin excitations. This peak is a 2D analog of the infinite-dimensional resonance peak of the dynamic mean-field approximation \cite{Georges} and the Abrikosov-Suhl resonance of the Anderson impurity model \cite{Hewson}. With further doping, the contribution of the upper Hubbard band decreases considerably and the DOS acquires some features of the weakly correlated spectrum (Figures~\ref{Fig2}(g) and (h)).

\begin{figure}[t]
\centering
\includegraphics[width=10 cm]{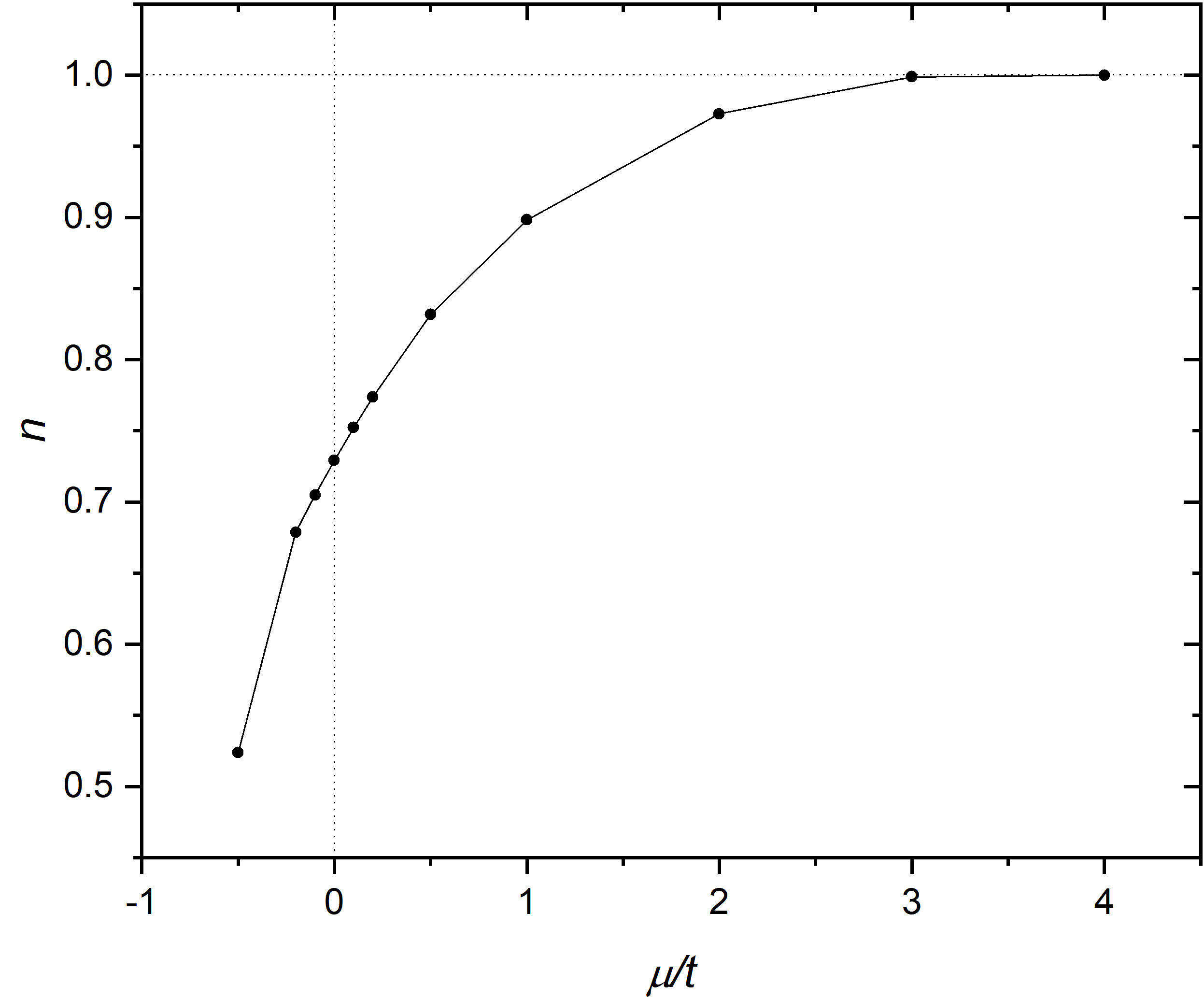}
\caption{The dependence $\bar{n}(\mu)$ for $U=8t$ and $T=0.32t$. The line connecting calculated points is guide for the eye.}\label{Fig3}
\end{figure}

The dependence $\bar{n}$ vs.\ $\mu$ for $U=8t$ and $T=0.32t$ is shown in Figure~\ref{Fig3}. It has the usual shape -- the concentration decreases monotonously with a fall in the chemical potential. However, already at this comparatively high temperature, some change in the dependence can be noticed for $\mu<0$.
The dependence changes radically with temperature lowering.
\begin{figure}[t]
\centering
\includegraphics[width=10 cm]{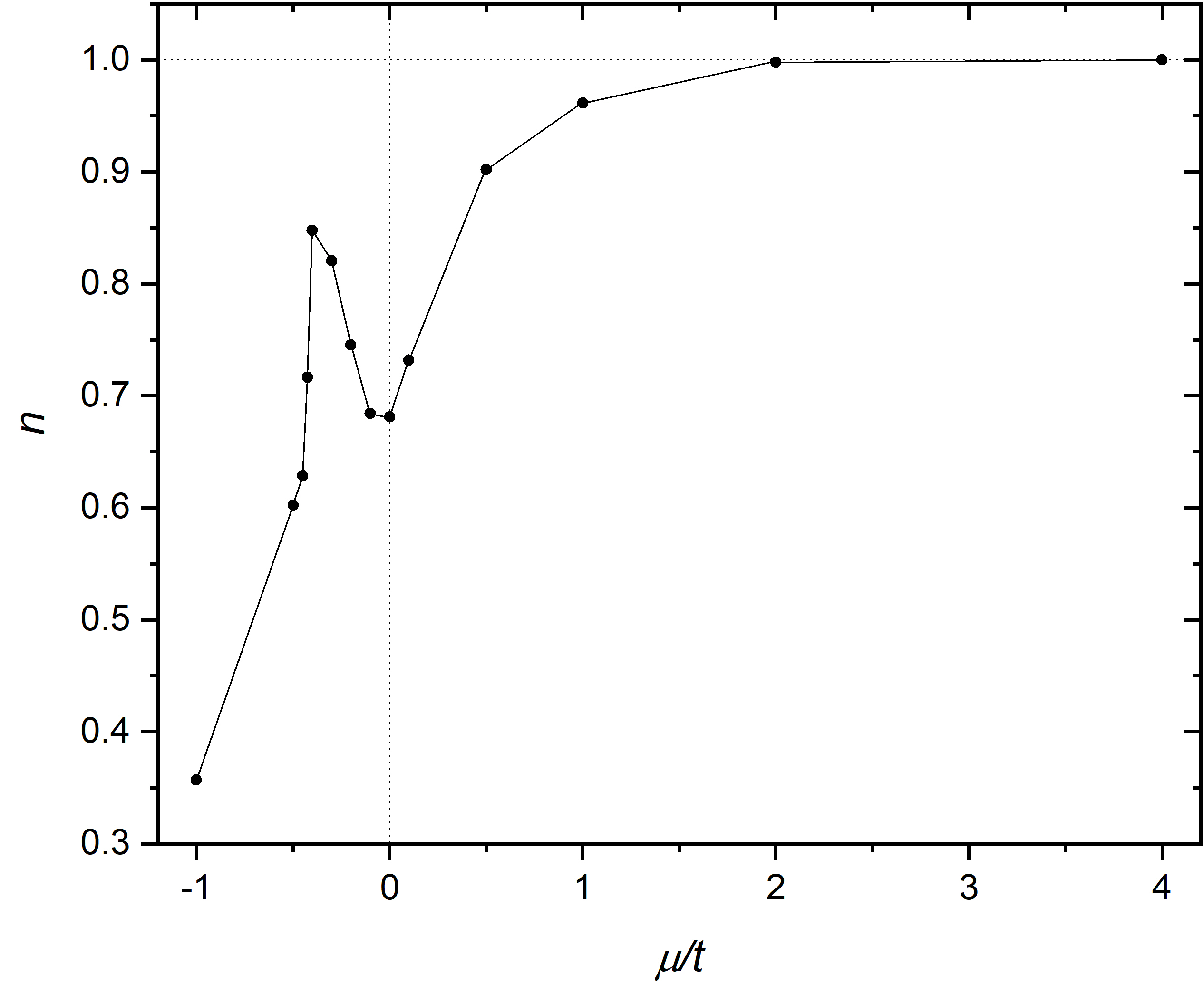}
\caption{Same as in Figure~\protect\ref{Fig3}, but for $T=0.13t$.}\label{Fig4}
\end{figure}
As seen in Figure~\ref{Fig4}, there appears a NEC region in the range $-0.4t\lesssim\mu\lesssim 0$. The mechanism of its occurrence can be understood from the expression for $C_1$ (\ref{C1}). Due to the discreteness of the energy levels of the Hubbard atom (\ref{Hubbard}), there are three regions of $\mu$, which correspond to its different fillings. In the first region, $\mu\ll -T$, $C_1(j)\approx 1/({\rm i}\omega_j+\mu)$, the atom is empty, and the solution of Hubbard model describes a gas of weakly correlated electrons. The second region, $T\ll\mu$ and $T\ll U-\mu$, $C_1(j)\approx(1/2)[1/({\rm i}\omega_j+\mu)+1/({\rm i}\omega_j+\mu-U)]$, relates to a lattice with predominantly singly occupied sites and strong electron correlations. The third region, $T\ll\mu-U$, $C_1(j)\approx 1/({\rm i}\omega_j+\mu-U)$ corresponds to a lattice with predominantly doubly occupied sites, in which single occupancies behave like a gas of weakly correlated species. It is clear that dependencies $\bar{n}(\mu)$ are different in these three situations that leads to NEC in transient regions $-T\lesssim\mu\lesssim T$ and $-T\lesssim U-\mu\lesssim T$. This simplified picture based on the Hubbard-I approximation is not modified considerably by charge and spin fluctuations, as can be seen in Figure~\ref{Fig4}.

With a further decrease of temperature, the NEC regions become narrower and the amplitude of changes in $\bar{n}$ gets apparently larger. Calculations for $T<0.1t$ need a larger number of frequency points and become too time and memory consuming for the used set of diagrams. However, for parameters corresponding to cuprates, $t\approx 0.2$~eV (see, e.g., \cite{Sherman18}) and $T=0.1t\approx 200$~K. Therefore, temperatures in the range of $0.01t-0.03t$ are closer to the usual experimental conditions. To attain this temperature region the irreducible part was simplified and approximated by the two lowest-order diagrams
\begin{equation}\label{K2}
K(j)=C_1(j)-\frac{T}{N}\sum_{{\bf k}j'\sigma'}C_2(j,\sigma;j,\sigma;j',\sigma';j',\sigma') \theta({\bf k},j').
\end{equation}
In this approximation, $K$ does not depend on momentum. The dependence $\bar{n}(\mu)$ calculated using (\ref{K2}) looks similar to those shown in Figures~\ref{Fig3} and \ref{Fig4} for the same temperatures. However, for lower temperatures, a new feature appears in this dependence -- a narrow region of bistability in the vicinity of the NEC.
\begin{figure}[t]
\centering
\includegraphics[width=10 cm]{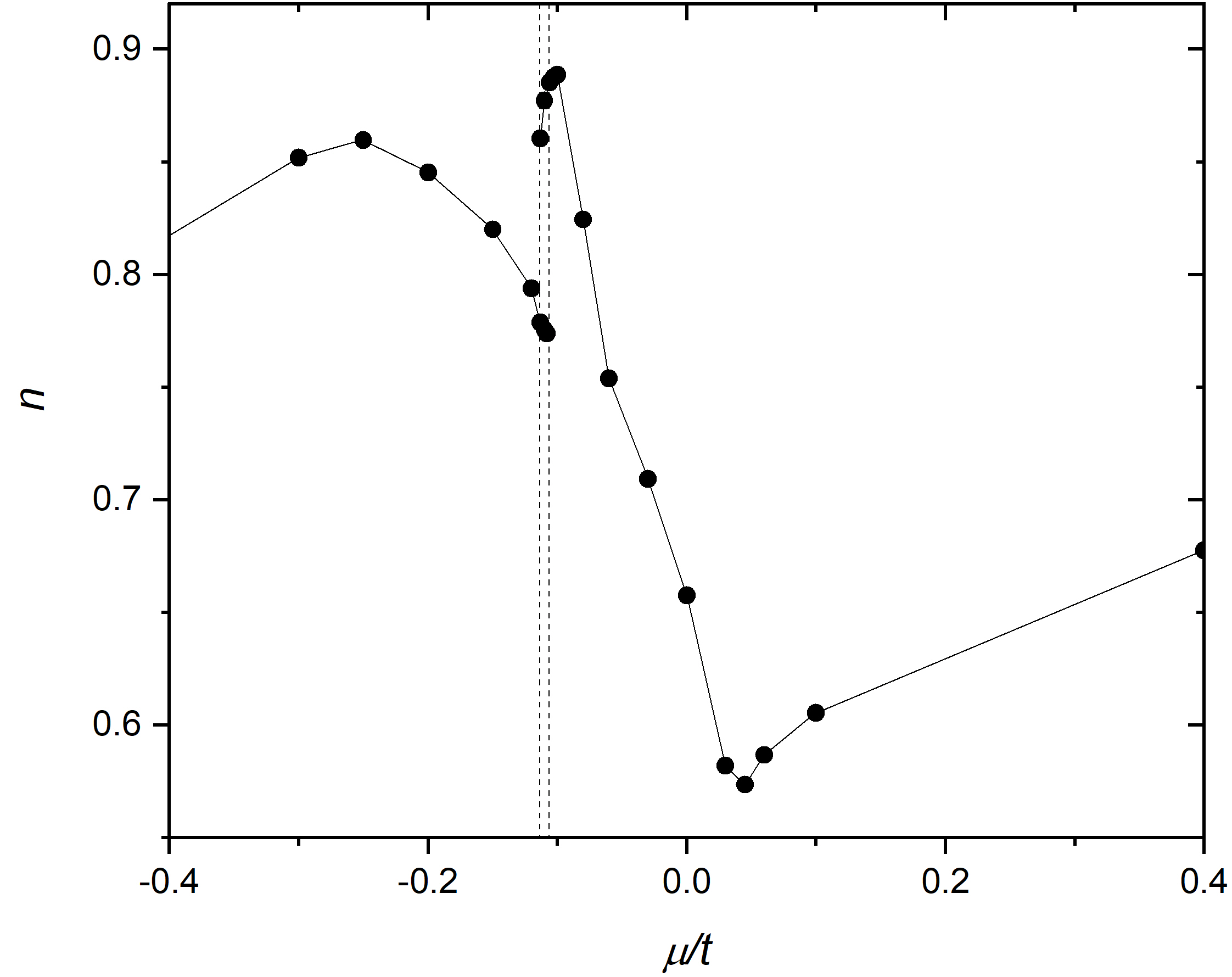}
\caption{The bistability region of the dependence $\bar{n}(\mu)$ obtained with the irreducible part (\protect\ref{K2}). $U=8t$, $T=0.03t$.}\label{Fig5}
\end{figure}
In this region, there are two values of the electron concentration for each chemical potential. Two branches of the dependence in Figure~\ref{Fig5} were obtained by taking a solution for some $\mu$ as a starting one in the iteration procedure, which was carried for a slightly larger or smaller chemical potential. For higher $T$ no bistability is observed.

The coexistence of two solutions with different $\bar{n}$ for a given value of the chemical potential points to the phase separation. Though this result has to be verified in a more elaborate approximation with a larger number of processes in $K$, it indicates the possibility of a purely electronic mechanism for the phase separation in crystals with strong electron correlations. The phase separation is ubiquitous in such compounds (see, e.g., \cite{Tranquada,Vershinin,Chu,Leshen}). However, underlying mechanisms of its creation are not well understood and strongly debated (see, e.g., \cite{Sherman08,Hizhnyakov,Sboychakov,Aichhorn,Otsuki}). In this connection, it is worth noting that in the range of chemical potentials $T\ll\mu$, $T\ll U-\mu$ no indication of the divergence in the charge susceptibility were found using the SCDT \cite{Sherman18,Sherman19b}. This fact excludes a purely electronic mechanism of the phase separation in this range of $\mu$ in the considered model. However, outside of the range, such a mechanism may exist, as seen in Figure~\ref{Fig5}.

\section{Concluding remarks}
In this work, the dependence of the hole concentration on the chemical potential was studied in the one-band repulsive Hubbard model on the two-dimensional square lattice in the case of strong correlations. Regions $\mu\leq 0$ and $\mu\geq U$ were of particular interest since due to the discreteness of the spectrum of the Hubbard atom and level crossing the negative charge compressibility was expected here. In the calculations of this work, the strong coupling diagram technique was used. Infinite sequences of diagrams describing interactions of electrons with spin and charge fluctuations were taken into account. This was done with the aim to prove the robustness of the NEC with respect to these fluctuations. The NEC regions were really found near $\mu=0$ and $U$ for low temperatures. For even lower temperatures, in the neighborhood of the NEC, there appear the bistability regions, in which two values of the electron concentration correspond to each value of the chemical potential. This is an indication of the phase separation in these regions. However, it should be noted that this result was obtained without regard for charge and spin fluctuations and needs in further verification. Due to more complex expressions for the second-order cumulants, a new technique was used for the solution of Bethe-Salpeter equations for charge and spin vertices. In the range $T\ll\mu$, $T\ll U-\mu$ obtained densities of states were similar to those derived in earlier works with the other method. For strong correlations, half-filling and moderate temperatures the DOS demonstrates the Mott gap near the Fermi level and reabsorption pseudogaps near $\pm U/2$. With doping, a pseudogap appears near the Fermi level at moderate temperatures. With the temperature decrease, the sharp peak develops at or in the nearest vicinity of the Fermi level. This peak is a manifestation of the bound states of correlated electrons and mobile spin excitations. It is the two-dimensional analog of the infinite-dimensional resonance peak of the dynamic mean-field approximation and the Abrikosov-Suhl resonance of the Anderson impurity model. For $\mu\lesssim 0$ and $\mu\gtrsim U$ the DOS acquires some features of a weakly correlated spectrum.

\printindex
\end{document}